\documentclass{article}
\usepackage{amssymb}

\usepackage{graphicx}
\usepackage{amsmath}

\input{tcilatex}

\begin{document}

\title{The Principle of Relativity and Modified Connection between the Mass of a
Body and its Velocity: a Case of $V>C.$}
\author{Elmir Dermendjiev}
\maketitle

\begin{abstract}
Following the basic idea expressed in [1], we assume that for any particle
or body with mass $M$ its own time $t$ depends on therelative change $\frac{%
\Delta M}{M}$ of that mass. Based on this assumption, one discusses possible
existence of a hypothetical particles and bodies with a velocity $V>C$
\end{abstract}

The main idea of paper [1] is based on the assumption that the flow of time $%
t$ which is associated with a given particle or body with mass $M$ is a
function of the relative change$\left( \frac{\Delta M}{M}\right) $ of its
mass:

\begin{equation}
dt=-\mu ^{-1}\frac{\Delta M}{M}
\end{equation}

The physical meaning of $\mu \;$is unknown: one does not know whether this
parameter is a function of coordinates, or a velocity $V$, or electrical
charge $Z$ etc. However, in [1]

it was noted that in spite of quite rough estimations of values of$\;\mu _{n}
$, $\mu _{w}$ and $\mu _{g}$ , a comparision of these parameters with
dimensionless constants C$_{s}$ , C$_{w}$ and C$_{g}$ for strong, weak and
gravitational interactions shows that both curves on Fig.1 of the paper [1]
are similar. That important observation was a serious reason to try-as much
as possible-to clarify physical meaning of $\mu \;$-parameters. Besides
that, below one discusses these conditions, which provide the velocity of a
particle or body to be of $V>C$. To get more information about the $\mu \;$%
-parameters, let us assume that at the front of moving particle a layer of \
matter with higher density arises. Since one does not know neither the
nature of assumed phenomenon nor any details of it, a simple model
consisting of a potential barrier with a height of $E_{b}$ and a thickness
of $\Delta x\;$can be applied. This allows the value of$\;E_{b}$ to be
roughly estimated. The probability $D\;$such barrier to be passed is given
by well known equation: 
\begin{equation}
D\thickapprox \exp \left( -\frac{2}{h}\sqrt{2M(E_{b}-E)}\Delta x\right) 
\end{equation}

Here $E$ is the energy of moving particle, which was defined in [1] as
follow:
\begin{equation}
E=\frac{\mu ^{2}}{2}(\Delta S)^{2}\left( \frac{\Delta M}{M}\right) ^{-2}
\end{equation}

If $E=E_{b}$ one gets $D=1$ and the height $E_{b}$ of the barrier can be
roughly

estimated. By using the values of $\Delta S\thickapprox 10^{-13}cm$ and of $%
\mu _{n}\thickapprox $10$^{23}\sec ^{-1}$

[1] one finds that $E_{b}\thickapprox 10^{20}$ $M\left( \frac{\Delta M}{M}%
\right) ^{-2}$. Assuming that $\left( \frac{\Delta M}{M}\right)
\longrightarrow 1$,

one can get an estimation of the minimum heigt:$E_{b}^{\min }$ $\thickapprox
10^{20}$ $M$. The same

approach can be used to estimate the value of $E_{b}(\beta )$ for $\beta $
-decay of a neutron. Using a value of$\;\mu _{w}\thickapprox 10^{-6}\sec
^{-1}$ [1] and of $\Delta S\thickapprox 10^{-13}$ $cm$ one obtains that $%
E_{b}(\beta )\thickapprox 10^{-38}$ $M\left( \frac{\Delta M}{M}\right) ^{-2}$%
.

\qquad As a next step one needs to define a value of $M$ for the electron

antineutrino, which is unknown. To make our estimation more conservative one

can assume that this mass lays in a quite wide range of $1eV<M<100eV$. In
addition, the value of $E_{b}(\beta )$ has to be comparable with the
difference between the neutron and proton masses, which is of $\thickapprox
1,3MeV$. This means that the value of $\left( \frac{\Delta M}{M}\right) ^{-2}
$. should be of $\thickapprox 10^{42}-$ $10^{44}$. Thus, the result leads us
to unexpected conclusion: since $\left( \frac{\Delta M}{M}\right) _{\nu
}\thickapprox 10^{-21}-$ $10^{-22}$, one should accept that the life-time of
the electron antineutrino could be $\thickapprox $ $10^{4}$times larger than
the life-time of Universe!

It is interesting to note that if $V<C$, where $C$- is the velocityof light,
then the value of$\;\mu $ is limited [1]:
\begin{equation}
\mu <\frac{\Delta M}{M}C(\Delta S)^{-1}
\end{equation}

On the other hand, since in [1] there are no any limitations the value of $%
V\;$to be larger than $C$, i.e.$V>C$, one can try to determine the
conditions that provide this case. The$\;\mu $-parameter has to satisfy the
following equation:
\begin{equation}
\mu >\frac{\Delta M}{M}C(\Delta S)^{-1}
\end{equation}

It is curios the values of $\mu _{n}(V>C)\;$ and $\mu _{w}(V>C)$ to be
estimated. 

One assumes that for strong interaction$\;\Delta S\thickapprox 10^{-13}cm$
and $\frac{\Delta M}{M}\leq 1$. The minimum estimated value of$\;\mu
_{n}(V>C)\thickapprox 10^{23}\sec ^{-1}$. Comparing both values of $\mu
_{n}(V<C)$ and $\mu _{n}(V>C)\;$one can find that the difference between
them is less than one order of magnitude.Similarly, one can get an
estimation of the value of $\mu _{w}(V>C)\thickapprox (10-10^{2})\sec ^{-1}$
for weak interaction:. Thus, we find the following ratios:
\begin{equation}
\frac{\mu _{n}(V>C)}{\mu _{n}(V<C)}\thickapprox 1
\end{equation}

\begin{equation}
\frac{\mu _{w}(V>C)}{\mu _{w}(V<C)}\thickapprox 10^{8}
\end{equation}

Considering these quite curious ratios one can express some very
interesting, but totally speculative thoughts. Indeed, since the values of
both$\;\mu _{n}$-parameters are almost the same, is it possible that in some
nuclear reactions an extremely small number of particles would penetrate
trough such high potential barrier via a tunneling and would have $V>C$?

On the other hand, since the above assumed potential barrier for $\beta $
-decay is much lower compare to strong interacting particle reactions, one
should expect higher yield of neutrinos with $V>C$.

\bigskip 

References

[1]. Elmir Dermendjiev, '' A Modified connection between the mass and time
in the principle of relativity'', preprint nucl-th 0011028 (08. November
2000)

\end{document}